\documentclass [11pt, letterpaper]{article}
\usepackage{fullpage}
\usepackage{amsmath}
\usepackage{amssymb}
\usepackage{graphicx}
\usepackage{mathrsfs}
\usepackage{wrapfig}
\usepackage{boxedminipage}
\usepackage{epsfig}
\usepackage{setspace}
\usepackage{subfigure}
\usepackage{float}
\usepackage{hyperref}
\usepackage{enumerate}
\usepackage[pdftex]{color}
\newcommand{\reef}[1]{(\ref{#1})}

\begin{document}

\begin{flushright}
\phantom{{\tt arXiv:1207.????}}
\end{flushright}

\bigskip
\bigskip
\bigskip

\begin{center} {\Large \bf Holography,  Fractionalization }
  
  \bigskip

{\Large\bf  and }

\bigskip

{\Large\bf  Magnetic Fields}

\end{center}

\bigskip \bigskip \bigskip \bigskip

\centerline{\bf Tameem Albash, Clifford V. Johnson, Scott MacDonald}

\bigskip
\bigskip
\bigskip

  \centerline{\it Department of Physics and Astronomy }
\centerline{\it University of Southern California}
\centerline{\it Los Angeles, CA 90089-0484, U.S.A.}

\bigskip

\centerline{\small \tt talbash,  johnson1, smacdona,  [at] usc.edu}

\bigskip
\bigskip


\begin{abstract} 
\noindent 
Four dimensional gravity with a $U(1)$ gauge field, coupled to various fields in asymptotically anti--de Sitter spacetime, provides a rich arena for the holographic study of the strongly coupled $(2{+}1)$--dimensional dynamics of finite density matter charged under a global $U(1)$. As a first step in furthering the study of the properties of fractionalized and partially fractionalized degrees of freedom in the strongly coupled theory, we construct electron star solutions at zero temperature  in the presence of a background magnetic field.  We work in  Einstein--Maxwell--dilaton theory. In all cases we construct, the magnetic source is cloaked by an event horizon. A key ingredient of our solutions is our observation  that starting with the standard Landau level structure for the density of states, the electron star limits reduce the charge density and energy density to that of the free fermion result.  Using this result we construct three types of solution: One has a star in the infra--red with an electrically neutral horizon, another has a star that begins at an electrically charged event horizon, and another has the star begin a finite distance from an electrically charged horizon. 
\end{abstract}
\newpage \baselineskip=18pt \setcounter{footnote}{0}
%
\section{Introduction}

Aspects of the low energy physics of matter charged under a global $U(1)$ at finite density can be studied at strong coupling using the AdS/CFT correspondence and deformations thereof \cite{Maldacena:1997re,Gubser:1998bc,Witten:1998qj,Aharony:1999t,Witten:1998zw} (for recent reviews, see refs.~\cite{McGreevy:2009xe,Hartnoll:2009sz,Herzog:2009xv,doi:10.1146/annurev-conmatphys-020911-125141,Hartnoll:2011fn}). While it remains unclear just how far--reaching the tools of holography will be in helping understand  experimentally accessible physics of various condensed matter systems, it is already apparent that potentially powerful new ways of characterizing several classes of important behaviour may be emerging from the lines of research underway. The language is that of a dual gravitational system, which has  the utility that it is often very geometrical in character, while also being in terms of  quantities that are gauge invariant. The dual effective field theories are usually formulated perturbatively as gauge theories (or generalizations thereof) for gauge group of rank $N$, where $N$ is large. The natural gauge invariant variables to use at strong coupling are usually less easy to work with. It is in  this strong coupling regime where the gravitational language is most effective.

At finite charge density, there is a non--zero gauge field $A_t$ switched on in the gravitational (``bulk'') background. (The gauged $U(1)$ there is the global $U(1)$ of the dual theory, according to the usual dictionary\cite{Maldacena:1997re,Gubser:1998bc,Witten:1998qj}). A most natural circumstance is to have an event horizon present, with a Reissner--Nordstrom black hole sourcing the electric flux, as first explored in this context in refs.~\cite{Chamblin:1999tk,Chamblin:1999hg}. Already there is interesting physics to be learned from such systems, giving insights into such phenomena as the holographic Hall and Nernst effects and other transport properties \cite{Hartnoll:2007ai,Hartnoll:2007ih,Hartnoll:2007ip}, holographic Fermi surfaces \cite{Lee:2008xf,Liu:2009dm,Cubrovic:2009ye,Faulkner:2009wj}, and so forth (see for example the reviews in refs.~\cite{McGreevy:2009xe,Hartnoll:2009sz,Herzog:2009xv}), but more recent insights have shown a wider context into which such horizon--endowed charged spacetimes fit nicely: The electric flux can be sourced by electrically charged fermionic matter in the bulk of spacetime {\it instead} of a horizon, giving a kind of charged ``star", where the geometry in the infra--red (IR) ceases to be AdS$_2\times \mathbb{R}^2$ (we stay with infinite volume henceforth in this discussion), and becomes of Lifshitz form\cite{Hartnoll:2010gu}. The heuristic argument here is that \cite{Hartnoll:2010gu} the matter in the bulk effectively screens the electric field in the IR, and the equations of motion yield the Lifshitz geometry, known to be appropriate when there is a massive gauge field \cite{Kachru:2008yh}.  One supplementary way to think about this solution is that it is a result of the black hole becoming unstable due to pair creation induced by the electric field. The charges separate in the field, one escaping to infinity while the other falls into the horizon (for suitable particle mass). The black hole loses charge and energy and ultimately a gas of charged matter, the star, is all that is left.

Naturally, the question arises as to what these two different situations (horizon source {\it vs.} matter source) represent for the dual low energy physics. Since at large $N$ the entropy of a spacetime with horizon has an extra factor of some power of $N$ (typically $N^2$ for ordinary gauge theory) compared to the entropy  for spacetimes without, it is clear that the degrees of freedom in each case are organized very differently, as noted very early on in holographic studies in the context of thermally driven  confinement--deconfinement phase transitions (modeled by the Hawking--Page transition \cite{Hawking:1982dh}) \cite{Witten:1998zw}.

The observation in the present context is that the difference here is between fully fractionalized and fully unfractionalized (or ``mesonic") phases of the low energy finite density system\footnote{The idea that these holographic systems may capture the dynamics of fractionalized phases seems to have first begun to emerge in the work of ref.~\cite{Faulkner:2010tq} in their semi--holographic approach to the low energy physics. There, they used the term ``quasiunparticle" for the effective particles in the unfractionalized phase.  In work dedicated to addressing the issue, ref.~\cite{Sachdev:2010um} further elucidated the connection between holographic physics and fractionalized phases. See refs.~\cite{doi:10.1146/annurev-conmatphys-020911-125141,Hartnoll:2011fn} for a review of some of these ideas.}. This is of some considerable interest in condensed matter physics, since descriptions of charge and/or spin separation (where the charge or spin degrees of freedom of an electron may move separately) are relatively common for getting access to certain types of physics often associated with the low energy dynamics of lattice models. The idea (aspects of it are reviewed in ref.~\cite{Hartnoll:2011fn}) is that the~$N^2$ degrees of freedom accessible when the horizon is present are typical of the larger number of fractionalized variables available (the analogue of the electron being split into separate charge and spin degrees of freedom plus a Lagrange multiplier field that becomes a dynamical gauge field at low energy), while the fewer degrees of freedom of the non--horizon systems are more like the unfractionalized ``electron" (a composite particle in this picture).  This fractionalization leads to a violation of the Luttinger theorem, where the fractionalized variables carry the missing charge density \cite{Hartnoll:2010xj,Iqbal:2011in,Sachdev:2011ze,Huijse:2011hp}.

This rather compelling picture is certainly worth exploring, since we have a fully non--perturbative tool for getting access to fractionalization. One exploration at zero temperature (of interest to us in the current work, as we shall see) is ref.~\cite{Hartnoll:2011pp}, where it was shown how to tune an operator (that is relevant in the ultraviolet (UV)) in the system that allows for a mixture of  fully fractionalized and fully unfractionalized phases. This is dual to turning on a dilaton--like scalar field in the bulk theory. The dilaton scalar field couples to gravity, has a non--trivial potential, and due to a term of the form: $ - \exp(\frac{2}{\sqrt{3}}\Phi)F^2$, (where $F$ is the gauge field strength) it  effectively spatially modulates the coupling strength of the bulk gauge field, allowing for different configurations of matter and gravity to minimize the action for given asymptotic fields. If the dilaton diverges positively in the IR, the effective Maxwell coupling vanishes in the IR, allowing for a geometry with a horizon again, while still having the matter back--react enough on the bulk geometry to  support a star that carries some of the electric charge. This gives a mixture of fractionalized and partially fractionalized phases.  If the dilaton diverges negatively in the IR, the effective Maxwell coupling diverges there and we have that the favoured situation is the purely ``mesonic" case where there is no horizon  and only charged matter. The negatively diverging dilaton serves to enhance the Maxwell sector's tendency to destabilize a charged black hole due to pair production, forcing it to evaporate completely in favour of a charged star. 

It is worth noting that the same way that the Einstein--Maxwell--AdS system in various dimensions may be consistently uplifted to 10 and 11 dimensional supergravity  as combinations of spins in the compact directions (actually, using equal spins \cite{Chamblin:1999tk}), the systems with a dilaton sector can be uplifted, but now all the spins are equal except one \cite{Cvetic:1999rb}. This is worth noting that since this means that the Einstein--Maxwell--dilaton--AdS system's consistent uplift may allow the fate of the regions with diverging dilaton to be studied in M--theory.

In each case, the reduced gravity system is coupled to a gas of charged fermions in the bulk, represented by a fluid stress energy tensor with a given pressure and density, and an appropriate equation of state is input in order to characterize the system.  This mirrors the standard construction of neutron stars using the Tolman--Oppenheimer--Volkoff method \cite{PhysRev.55.374,PhysRev.55.364} which was first generalized to asymptotically AdS geometries in ref.~\cite{deBoer:2009wk}.

In this paper, we begin the exploration of the important situation of having a magnetic field present in the system.  We learn some very interesting lessons from this study. One of them is that we need a dilaton present to achieve unfractionalized phases of the physics. Since we are including no magnetically charged matter in our system, and in the limit we take there are no current sources of magnetic field, for a smooth gravitational dual it is intuitively clear that there must be a horizon to source the magnetic field. One can imagine therefore that there might be configurations where such a magnetic source is surrounded by a gas of matter that forms a star that sources all the electric flux. This would be the fully unfractionalized magnetic case, with the purely dyonic Reissner--Nordstrom black hole being the fractionalized counterpart. 

In fact, it turns out that the matter alone is unable to back--react enough on the magnetic geometry that we find to support a star. In other words, the magnetic field's presence stops the electric field from giving the local chemical potential (as seen by the charged bulk matter) the profile needed to support a star.  Put differently,  the magnetic field suppresses the pair--creation channel by which a black hole can leak its charge into a surrounding cloud of fermions to form a star. This is where the dilaton can come in. A negatively diverging dilaton in the IR will send the Maxwell coupling to infinity. This makes the pair--creation channel viable again, and allow for the system to seek stable solutions that have some of the electric flux sourced by fermions outside the horizon. 

In summary we find that in the limit we are working, when there is a magnetic field in the system there is necessarily an event horizon, and it is essential to have a dilaton present to not have all the electric flux sourced behind the horizon (at least for the zero temperature case we study here). Introducing the dilaton in the theory allows us to construct a mesonic phase as well as a partially fractionalized phase (the fully fractionalized phase is simply the already known dilaton--dyon solution).  

The work we report on in this paper is but a first step in characterizing these systems in the presence of magnetic field.  Our goal here is to exhibit the solutions we find\footnote{We emphasize that our work is both qualitatively and quantitatively different from ref.~\cite{Burikham:2012kn}.  Their work is in five dimensions, for a spherical star, and their TOV treatment involves an electrically neutral star.}.  We will postpone the analysis of the fluctuations of the solutions  to examine various transport properties of the system, and also leave for a later time the comparison of the actions of the various solutions for given magnetic field and chemical potential, which allow a full exploration of the phase diagram.

\section{Charged Ideal Fluid with a Magnetic Field}
%
\subsection{Gravity background}
We consider a simple  model of  Einstein--Maxwell theory coupled to a dilaton $\Phi$ and an ideal fluid source (akin to that of ref.~\cite{Hartnoll:2011pp}).  The Einstein--Maxwell sector has action given by:
\begin{equation}
S_{\rm EM} = \int d^4 x \sqrt{-G} \left[ \frac{1}{2 \kappa^2} \left( R - 2 \partial_\mu \Phi \partial^\mu \Phi - \frac{V(\Phi)}{L^2} \right) - \frac{Z(\Phi)}{4 e^2} F_{\mu \nu}F^{\mu \nu} \right] \ .
\end{equation}
The Einstein equations of motion are given by:
\begin{equation}
R_{\mu \nu} -\frac{1}{2} \left( R - 2 \partial_\lambda \Phi \partial^\lambda \Phi  -  \frac{V(\Phi)}{L^2} \right) G_{\mu \nu} - 2 \partial_\mu \Phi \partial_\nu \Phi= \kappa^2 \left( \frac{Z(\Phi)}{e^2} \left( F_{\mu \rho} F_{\nu}^{ \ \rho} - \frac{1}{4} G_{\mu \nu} F_{\lambda \rho} F^{\lambda \rho} \right) + T^{\mathrm{fluid}}_{\mu \nu} \right) \ ,
\end{equation}
with the following stress--energy tensor and external current for a perfect (charged) fluid (the action that gives rise to this stress energy tensor is given in subsection \ref{sec:Action}):
\begin{equation}
T^{\mathrm{fluid}}_{\mu \nu} = \frac{1}{L^2 \kappa^2}\left( \left( \tilde{P}(r) + \tilde{\rho}(r) \right) u_\mu u_\nu + \tilde{P}(r) G_{\mu \nu} \right) \ , \quad J_\mu =\frac{1}{e L^2 \kappa} \tilde{\sigma}(r) u_\mu \ .
\end{equation}
The Maxwell equations are given by:
\begin{equation}
\partial_\nu \left( \sqrt{-G} Z(\Phi) F^{\mu \nu} \right) = e^2 \sqrt{-G} J^{\mu} \ .
\end{equation}
We begin with the following metric ansatz:
\begin{equation}
ds^2 = L^2 \left( - f(r) dt^2 + a(r) d\vec{x}^2 + g(r) dr^2 \right) \ ,
\end{equation}
and we take $u_t =- \sqrt{-G_{tt}}$.  For the Maxwell field, we take as ansatz:
\begin{equation}
F_{r t} = \frac{e L}{\kappa} h'(r) \ , \quad F_{x y} = \frac{e L}{\kappa} \tilde{B} \ . 
\end{equation}
The resulting equations of motion can be reduced (after some work) to the following six equations:
\begin{eqnarray}
&\tilde{P}'(r) + \frac{f'(r)}{2f(r)} \left( \tilde{P}(r) + \tilde{\rho}(r) \right) - \tilde{\sigma}(r) \frac{h'(r)}{\sqrt{f(r)}}= 0 \ , & \label{eqt:Pressure} \\
&a''(r) - a'(r) \left(\frac{g'(r)}{2 g(r)}  + \frac{f'(r)}{2 f(r)} + \frac{a'(r)}{2 a(r)} \right) + a(r) \left( g(r) \left( \tilde{P}(r) + \tilde{\rho}(r) \right) +2 \Phi'(r)^2\right)= 0 \ , & \label{eqt:a}\\
&\frac{f''(r)}{f(r)} - \frac{f'(r)}{f(r)} \left( \frac{g'(r)}{2 g(r)} + \frac{f'(r)}{2 f(r)} - \frac{2 a'(r)}{a(r)} \right) +  \left( \frac{a'(r)^2}{2 a(r)^2} - 2 \Phi'(r)^2 - g(r) \left( 5 \tilde{P}(r) + \tilde{\rho}(r) - 2 V(\Phi) \right) \right) = 0  \ , & \label{eqt:f} \\
&\Phi'(r)^2 + g(r) \left(- \frac{Z(\Phi) \tilde{B}^2}{2 a(r)^2} + \left( \tilde{P}(r) - \frac{1}{2} V(\Phi)\right) \right) - \frac{a'(r)^2}{4 a(r)^2} - \frac{a'(r) f'(r)}{2a(r) f(r)} - \frac{Z(\Phi) h'(r)^2}{2 f(r)} = 0  \ , & \label{eqt:constraint} \\
&h''(r) - h'(r) \left( \frac{g'(r)}{2 g(r)} + \frac{f'(r)}{2 f(r)} - \frac{ a'(r)}{a(r)} - \frac{Z'(\Phi) \Phi'(r)}{Z(\Phi)} \right) -  \frac{\sqrt{f(r)} g(r)}{Z(\Phi)} \tilde{\sigma}(r) = 0  \ , & \\
&\Phi''(r) + \Phi'(r) \left( \frac{f'(r)}{2 f(r)} - \frac{g'(r)}{2 g(r)} + \frac{a'(r)}{a(r)} \right) - \frac{Z'(\Phi)}{4} \left( \frac{g(r) \tilde{B}^2}{a(r)^2} - \frac{h'(r)^2}{f(r)} \right) - \frac{g(r) V'(\Phi)}{4} = 0 \ . &
\end{eqnarray}
For the charge density $\tilde{\sigma}(r)$ and the energy density $\tilde{\rho}(r)$, we show in subsection \ref{sec:DoS} that using the density of states of charged quanta in a magnetic field results in   the free fermion gas result when using the appropriate electron star limits:
\begin{eqnarray}
&\tilde{\sigma}(r) = \frac{1}{3} \tilde{\beta} \left( \tilde{\mu}(r)^2 - \tilde{m}^2 \right)^{3/2}\ , & \nonumber \\
& \tilde{\rho}(r) = \frac{1}{8} \tilde{\beta} \left( \tilde{\mu}(r) \sqrt{ \tilde{\mu}(r)^2 - \tilde{m}^2 } \left( 2 \tilde{\mu}(r)^2 - \tilde{m}^2 \right) + \tilde{m}^4 \ln \left( \frac{\tilde{m}}{\tilde{\mu}(r) + \sqrt{ \tilde{\mu}(r)^2 - \tilde{m}^2}} \right) \right) \ , &
\end{eqnarray}
and equation \reef{eqt:Pressure} is satisfied with the following equation for the pressure:
\begin{equation}
\tilde{P}(r) = - \tilde{\rho}(r) + \tilde{\mu}(r) \tilde{\sigma}(r) \ .
\end{equation}
where we have defined $\tilde{\mu}(r) = h(r)/\sqrt{f(r)}$ as the local chemical potential.  Here, $\tilde{m}=\kappa m/e$, where $m$ is the fermion mass.
Equations \reef{eqt:a}, \reef{eqt:f}, and \reef{eqt:constraint} are derived from the Einstein equations of motion, and only two of them are dynamical equations; the remaining equation is a constraint.  For concreteness, let us take from this point forward:
\begin{equation}
V(\Phi) = -6 \cosh(2 \Phi/\sqrt{3}) \ , \quad Z(\Phi) = e^{2 \Phi/\sqrt{3}} \ .
\end{equation}
This choice matches the choice in ref.~\cite{Gubser:2009qt} that give rise to three--equal--charge dilatonic black holes in four dimensions.  We want our solutions to asymptote to AdS$_4$ so we will require that the UV behavior of our fields is given by:
\begin{equation}
g(r) = \frac{1}{r^2}  \ , \quad f(r) = \frac{1}{r^2} \ , \quad a(r) = \frac{1}{r^2} \ ,
\end{equation}
which in turn gives the following UV behavior for the remaining fields:
\begin{equation}
\Phi(r) = r \phi_1 + r^2 \phi_2 \ , \quad h(r) = \mu - \rho \ r \ ,
\end{equation}
where $\phi_1$ is proportional to the source of the operator dual to the dilaton, $\phi_2$ to the vev of the same operator, $\mu$ the chemical potential in the dual field theory, and $\rho$ the charge density.

\subsection{Density of States for $(3{+}1)$--Dimensional Fermions in a Magnetic Field} \label{sec:DoS}
The appropriate choice for the density of states should be that of the 3 (spatial) dimensional Landau levels.  For a fermion of  mass $m$ with charge $q$ in the $n$--th Landau level with  momentum $k$, the energy in magnetic field $B$ is given by ($ c= \hbar = 1$):
\begin{equation}
E_n(k) = \sqrt{2 q B (n+1) +  k^2 + m^2} \ ,
\end{equation}
where for simplicity, we are assuming that $q B > 0$.  At level $n$ and momentum $k$, the number of states is given by:
\begin{equation}
N_n(k) = g_s  \left( \frac{ k L_r}{2 \pi} \right) \left( 2 q B \frac{L_x L_y}{4 \pi} \right) \ , 
\end{equation}
where $g_s$ is the spin degeneracy ($=2$ for spin 1/2).  In turn, at a given energy $E$, the number of states is given by:
\begin{eqnarray}
N(E) &=& g_s \sum_{n=0}^\infty \frac{ L_x L_y L_r}{8 \pi^2} 2 q B \sqrt{E^2 - m^2 - 2 q B (n+1)} \, \Theta\!\left( E - \sqrt{m^2 + 2 q B(n+1)} \right) \nonumber \\
&=& V \frac{\beta}{2} 2 q B \sum_{n} \sqrt{E^2 - m^2 - 2 q B (n+1)}\, \Theta\!\left( E - \sqrt{m^2 + 2 q B(n+1)} \right) \ ,
\end{eqnarray}
where we have defined a dimensionless constant of proportionality which is $O(1)$.  Therefore, we can write the density of states as:
\begin{eqnarray}
g(E) &=&\frac{\beta}{2} 2 q B \sum_{n=0}^{\infty} \left[ \frac{ E}{\sqrt{E^2 - m^2 -2 q  B(n+1)}} \, \Theta\!\left( E - \sqrt{m^2 + 2 q B(n+1)} \right)  \right. \nonumber \\
&& \left.+ \sqrt{E^2 - m^2 - 2 q B (n+1)}\, \delta\! \left( E - \sqrt{m^2 + 2 q B(n+1)} \right) \right] \ .
\end{eqnarray}
The second term will always give us zero contribution for the terms we are interested in, so we will drop it.
The energy density $\rho$ and charge density $\sigma$ are given by:
\begin{equation}
\sigma = \int_0^\mu d E \  g(E) \ , \quad \rho = \int_0^\mu d E \  E g(E) \ ,
\end{equation}
where $\mu$ is the chemical potential.  To proceed, we identify the local magnetic field in our gravity background with $q B$ and the local chemical potential with $\mu$:
\begin{equation}
q B = \frac{F_{x y}}{L^2 A(r)} = \frac{e}{L \kappa} \frac{\tilde{B}}{A(r)} \ , \quad \mu = \frac{A_t}{L \sqrt{ f(r)}} = \frac{e}{\kappa} \frac{h(r)}{\sqrt{f(r)}} \equiv \frac{e}{k} \tilde{\mu} \ .
\end{equation}
For conciseness, let us define:
\begin{equation}
\tilde{q} \equiv \frac{\kappa}{L e^2} \frac{2 e}{A(r)} \ .
\end{equation}
In terms of dimensionless variables $\tilde{\sigma}(r) \equiv e L^2 \kappa \sigma(r)$ and $\tilde{\rho}(r) \equiv L^2 \kappa^2 \rho(r)$, we now have:
\begin{eqnarray} \label{eqt:sigma_rho}
\tilde{\sigma}(r) &=&\frac{\tilde{\beta}}{2} 2 \tilde{q} \tilde{B} \sum_{n=0}^\infty \Theta \left(\tilde{\mu}^2 -  \tilde{m}^2 -2 \tilde{q} \tilde{B}(n+1) \right) \int_{\sqrt{\tilde{m}^2 +2 \tilde{q} \tilde{B}(n+1)}}^{\tilde{\mu}} \frac{d \tilde{E}  \ \tilde{E}}{\sqrt{\tilde{E}^2 - \tilde{m}^2 -2 \tilde{q} \tilde{B}(n+1) }} \ , \nonumber  \\
\tilde{\rho}(r) &=&\frac{\tilde{\beta}}{2} 2 \tilde{q} \tilde{B} \sum_{n=0}^\infty \Theta \left(\tilde{\mu}^2 -  \tilde{m}^2 -2 \tilde{q} \tilde{B}(n+1) \right) \int_{\sqrt{\tilde{m}^2 +2 \tilde{q} \tilde{B}(n+1)}}^{\tilde{\mu}} \frac{d \tilde{E}  \ \tilde{E}^2}{\sqrt{\tilde{E}^2 - \tilde{m}^2 - 2\tilde{q} \tilde{B}(n+1)}}  \ .
\end{eqnarray}
where $\tilde{\beta} = e^4 L^2 \beta /\kappa^2$ and $\tilde{m} =   \kappa m / e$.  To be in the classical gravity regime, we must have $\kappa/L \ll 1$.  To be able to use the flat space physics, we must have a large density relative to the curvature scale of the geometry, we must have $\sigma L^3 \sim L \sigma(r) / e \kappa \gg 1$.  To have $\tilde{\beta} \sim O(1)$, we must have that $e^2 L  / \kappa \sim O(1)$.  Therefore by our previous requirements, we must have $e^2 \sim \kappa / L$.  These are the requirements from ref.~\cite{Hartnoll:2010gu}.  With these requirements, we have the following result:
\begin{equation}
2 \tilde{q} \tilde{B} \ll 1\ .
\end{equation}
We can perform the integrals explicitly.  We have that:
\begin{eqnarray}
\int_{\sqrt{\tilde{m}^2 +2\tilde{q} \tilde{B}(n+1)}}^{\tilde{\mu}} \frac{d \tilde{E}  \ \tilde{E}}{\sqrt{\tilde{E}^2 - \tilde{m}^2 -2 \tilde{q} \tilde{B}(n+1) }} &=& \sqrt{ \tilde{\mu}^2 - \tilde{m}^2 - 2 \tilde{q} \tilde{B} (n+1)} \nonumber \\
\int_{\sqrt{\tilde{m}^2 +2\tilde{q} \tilde{B}(n+1)}}^{\tilde{\mu}} \frac{d \tilde{E}  \ \tilde{E}^2}{\sqrt{\tilde{E}^2 - \tilde{m}^2 -2 \tilde{q} \tilde{B}(n+1) }} &=& \frac{1}{2} \left[ \tilde{\mu} \sqrt{\tilde{\mu}^2 - \tilde{m}^2 - 2 \tilde{q} \tilde{B} (n+1)}   \right. \nonumber \\
&&\hskip-2.25cm \left. - \left( \tilde{m}^2 + 2 \tilde{q} \tilde{B} (n+1) \right) \ln \left( \frac{ \sqrt{ \tilde{m}^2 + 2 \tilde{q}  \tilde{B} (n+1)}}{\tilde{\mu} + \sqrt{ \tilde{m}^2 + 2  \tilde{q} \tilde{B} (n+1)}} \right) \right] \ .
\end{eqnarray}
Furthermore, we see that the spacing between Landau levels is extremely small, and to zeroth order in $2 \tilde{q} \tilde{B}$ we can approximate the sum by an integral (Euler--Maclaurin):
\begin{equation}
2 \tilde{q} \tilde{B}    \sum_{n=0}^\infty \Theta \left(\tilde{\mu}^2 - m^2 - 2 \tilde{q} \tilde{B}(n+1)  \right) F \left(2 \tilde{q} \tilde{B} (n+1) \right) \approx \int_{0}^{\tilde{\mu}^2 - \tilde{m}^2}\!\! F(x) d x + O \left(2 \tilde{q} \tilde{B} \right) \ .
\end{equation}
We can perform this final integration to give us our final result for the charge density and energy density:
\begin{eqnarray}
\tilde{\sigma} &=& \frac{1}{3} \tilde{\beta} \left( \tilde{\mu}^2 - \tilde{m}^2 \right)^{3/2} + O \left(2 \tilde{q} \tilde{B} \right) \nonumber \\
\tilde{\rho} &=& \frac{1}{8} \tilde{\beta} \left( \tilde{\mu} \sqrt{ \tilde{\mu}^2 - \tilde{m}^2 } \left( 2 \tilde{\mu}^2 - \tilde{m}^2 \right) + \tilde{m}^4 \ln \left( \frac{\tilde{m}}{\tilde{\mu} + \sqrt{ \tilde{\mu}^2 - \tilde{m}^2}} \right) \right) + O \left(2 \tilde{q} \tilde{B} \right) \ .
\end{eqnarray}
Interestingly enough, the zeroth order results are exactly the free fermion gas results in the absence of a magnetic field.  Since the pressure equation (eqn. \reef{eqt:Pressure}) is the same as that in the free fermion in the absence of a magnetic field case, the ansatz:
\begin{equation}
\tilde{P} = - \tilde{\rho} + \tilde{\mu} \tilde{\sigma} \ ,
\end{equation}
trivially satisfies that equation.
%
\subsection{Action Calculation} \label{sec:Action}
%
We discuss in detail an action that recovers the equations of motion used for our background.  Our calculation generalizes the calculation performed in ref.~\cite{Hartnoll:2010gu} to the case with a magnetic field.  We consider an action for the fluid along the lines of refs.~\cite{Brown:1992kc,0264-9381-7-10-008,PhysRevD.31.1854} given by:
\begin{equation}
\mathcal{L}_{\mathrm{fluid}} = \sqrt{-G} \left(  -\rho(\sigma) + \sigma u^\mu \left( \partial_\mu \phi + A_\mu + \alpha \partial_\mu \beta \right) + \lambda \left( u^\mu u_\mu +1 \right) \right) \ ,
\end{equation}
where $\phi$ is a Clebsch potential variable, $(\alpha,\beta)$ are potential variables, and $\lambda$ is a Lagrange multiplier.  The equations of motion from this part of the action give:
\begin{eqnarray}
\delta \sigma: && - \rho' (\sigma) + u^\mu \left( \partial_\mu \phi + A_\mu + \alpha \partial_\mu \beta \right) = 0  \ , \\
\delta u_\mu: &&  G^{\mu \nu} \left( \sigma \left(\partial_\nu \phi + A_\nu + \alpha \partial_\nu \beta \right) + 2 \lambda u_\nu \right)  = 0 \ , \\
\delta \lambda: && u_\mu u^{\mu} = -1  \ , \\
\delta \phi: && \partial_\mu \left( \sqrt{-G} G^{\mu \nu} \sigma u_\nu \right) = 0 \ , \\
\delta \alpha: && u^\mu \partial_\mu \beta = 0 \ , \\
\delta \beta: && \partial_\mu \left( \sqrt{-G} \sigma u^\mu \alpha \right) = 0 \ . 
\end{eqnarray}
We define the local chemical potential $\mu \equiv \rho'(\sigma)$ such that:
\begin{equation}
\mu = u^{\mu} \left( \partial_\mu \phi + A_{\mu} + \alpha \partial_\mu \beta  \right)  \ .
\end{equation}
From the equation of the fluctuation of $u_\mu$, we can multiply through by $u_\mu$ to fix the Lagrange multiplier:
\begin{equation}
\lambda = \frac{1}{2} \sigma \mu = \frac{1}{2}  \left( P + \rho \right) \ ,
\end{equation}
where we have used the thermodynamic relation $P = - \rho + \mu \sigma$.  Now we recall that we wish to fix $u_t = - \sqrt{-G_{tt}}, u_x = u_y = 0, \mu = A_t/(- u_t)$ and choose a gauge where $A_x = 0, A_y = e L \tilde{B} x / \kappa$.  Since our metric ansatz has $G^{xx} = G^{yy}$, we can satisfy the equation for the fluctuation of $u_\mu$ (and all the equations) with:
\begin{equation}
\phi = -  \frac{e L }{\kappa} \tilde{B}  x y  \ , \quad \alpha =  \frac{e L }{\kappa}  \tilde{B} y \ , \quad \beta = x \ .
\end{equation}
with these choices, we note that:
\begin{equation}
\frac{\delta \mathcal{L}_{\mathrm{fluid}}}{\delta G^{\mu \nu}} = \sqrt{-G} \left(- \frac{1}{2} G_{\mu \nu} \left( - \rho + \mu \sigma \right) -\frac{1}{2} \delta_\mu^t \delta_\nu^t \mu \sigma u_t u_t \right) \ ,
\end{equation}
which for us recovers the (on--shell) fluid energy-momentum tensor using in our equations of motion:
\begin{equation}
T_{\mu \nu}^{\mathrm{fluid}} = \frac{-2}{\sqrt{-G}} \frac{\delta \mathcal{L}_{\mathrm{fluid}}}{\delta G^{\mu \nu}} = \left( P + \rho \right) u_\mu u_\nu + P G_{\mu \nu} = \frac{1}{L^2 \kappa^2} \left( \left( \tilde{P} + \tilde{\rho} \right) u_\mu u_\nu + \tilde{P} G_{\mu \nu} \right)  \ .
\end{equation}

\section{The Role of the Dilaton}
%
We first remark on why Einstein--Maxwell theory alone ({\it i.e.}' without a dilaton) is not sufficient for studying backgrounds with an electric star in the presence of a magnetic field.  Let us consider the case where the dilaton is absent in the theory.  We find the following series expansion in the IR ($r \to \infty$):
\begin{equation} \label{eqt:PureDyon}
a(r) = \sum_{n=0}^{\infty} \frac{a_n}{r^n} \ ,  \quad f(r) = \frac{1}{r^2} \sum_{n=0}^{\infty} \frac{f_n}{r^n} \ , \quad g(r) = \frac{1}{r^2} \sum_{n=0}^{\infty} \frac{g_n}{r^n} \ , \quad h(r) = \frac{1}{r} \sum_{n=0}^{\infty} \frac{h_n}{r^n} \ , 
\end{equation}
where the leading coefficients fixed by the equations of motion are given by:
\begin{equation}
g_0 = \frac{1}{6} \ , \quad \frac{h_0}{\sqrt{f_0}} =  \sqrt{1- \frac{\tilde{B}^2}{6 a_0^2}} \ ,
\end{equation}
where we have picked the positively charged horizon (sign of $h_0$) without loss of generality.  This is of course the well established near horizon geometry of the extremal dyonic black hole in AdS$_4$.  We find that all the coefficients of $f(r)$ are undetermined by the equations of motion, as well as $a_0$ and $a_1$.  The value of $a_0$ is fixed by requiring that $r^2 a(r)$ be equal to one in the UV, so it is the ratio $a_1/a_0$ that is the free parameter.  We note that the near horizon behavior of the local chemical potential is:
\begin{equation} \label{eqt:PureDyonMu}
\frac{h(r)}{\sqrt{f(r)}} = \frac{h_0}{\sqrt{f_0}} \left(1 - \frac{1}{r} \frac{a_1}{a_0} f_0 + O\left(r^{-2} \right) \right) \ .
\end{equation}
This chemical potential \emph{decreases} (this is because only for $a_1 > 0$ do we get an asymptotically AdS$_4$ solution) from its IR value.  This will be an important point for what comes next.

Let us now consider the case where we turn on the source and look for a solution where we have a star in the IR.  We find that an identical expansion as equation \reef{eqt:PureDyon} works, but the leading coefficients are given by:
\begin{equation}
g_0 = \frac{1}{6} \ , \quad \frac{h_0}{\sqrt{f_0}} = \tilde{m} \ , \quad \tilde{B}^2 = 6 a_0^2 \left( 1- \tilde{m}^2 \right) \ .
\end{equation}
However, the situation is more grave than this; the only solution allowed is:
\begin{equation}
\frac{h(r)}{\sqrt{f(r)}} = \tilde{m} \ , \quad a(r) = a_0 \ ,
\end{equation}
which is clearly not an asymptotically AdS$_4$ solution (one can attempt to generalize the IR expansion to having $f(r) \propto r^{-2z}$ and $g(r) \propto r^{-z}$ but the results remain the same up to factors of $z$).  For $f(r) = 1/r^2$, the solution is simply the near horizon geometry of a dyon where the electric charge has been set by $\tilde{m}$.  Furthermore, since the chemical potential is equal to $\tilde{m}$ everywhere, there is no star, so we cannot construct a solution with a star in the IR.

We may consider building a solution where there is no star in the IR but instead exists a finite distance away from the horizon.  Therefore, we would have the same IR expansion as equation \reef{eqt:PureDyon}, and hope that we are able to populate the star at some finite distance $r < \infty$.  This would require $\tilde{m} > h_0/\sqrt{f_0}$. but, as pointed out in equation \reef{eqt:PureDyonMu}, we find that the chemical potential decreases as we move away from the horizon.  In fact, (as we know already from the exact solution) the chemical potential monotonically decreases from the horizon to the UV, so a star can never form.  Therefore, we find that we cannot support a magnetic star in the simple Einstein--Maxwell setup.

One way forward is to introduce the dilaton in the theory, which  changes the effective coupling strength of the Maxwell field. As mentioned in the introduction, a dilaton was first used in this context in ref.~\cite{Hartnoll:2011pp}. Introducing the dilaton is dual to  using a relevant operator (in the UV) to induce the theory to flow to different IR phases.  Three different phases were identified: a ``mesonic'' phase, where all the electric charge is \emph{not} behind the horizon, a partially ``fractionalized'' phase where \emph{a fraction} of the charge is behind the horizon, and a fully fractionalized phase, where all the charge is behind the horizon.  We will use this nomenclature to label our solutions.

The dilaton's behavior in the IR naturally provides us a classification scheme for our solutions.  It is useful to review this before introducing the star.  The dilaton can diverge positively or negatively in the IR, giving rise to a purely electric or purely magnetic horizon, or it can take a finite value, giving a dyon solution.  Since we are only interested in geometries with a magnetic field, we will ignore the purely electric case in what follows.  We briefly review the IR asymptotics of the two magnetic cases.
The purely magnetic case has an IR ($r \to \infty$) expansion for the fields:
\begin{eqnarray}
f(r) &=& \frac{1}{r^2} \left( \sum_{n=0}^{\infty} \frac{f_n}{r^{4n/3}} \right)  \  , \quad g(r) = \frac{1}{r^{8/3}} \left(\sum_{n=0}^{\infty} \frac{g_n}{r^{4n/3}} \right) \ ,\quad  a(r) =  \frac{1}{r^{2/3}} \left(\sum_{n=0}^{\infty} \frac{a_n}{r^{4n/3}} \right) \ , \nonumber \\
\Phi(r) &=& - \frac{\sqrt{3}}{3} \ln r + \left(\sum_{n=0}^{\infty} \frac{\Phi_n}{r^{4n/3}} \right)  \ , \quad h(r) = 0 \ , \quad \tilde{B}^2 = \frac{3}{2} e^{-4 \Phi_0 / \sqrt{3}} a_0^2 \ .
\end{eqnarray}
The coefficients $f_n$ are completely undetermined by the equations of motion as well as $\Phi_0$ and $a_0$ (note that $a_0$ is not really a free parameter since it determines what the coefficient for $d \vec{x}^2$ is in the UV which needs to be 1).
Therefore, we see that there are two free parameters in the IR, $\Phi_0$ and the function $f(r)$ (subject to obeying the right asymptotics in the UV and IR).  This translates to the freedom of choosing a particular magnetic field and source for the dilaton in the dual field theory.  Finally, we note that the behavior of the dilaton in the IR means that the Maxwell coupling is diverging, which means that quantum loop effects would become important near the horizon.

The dyon solution has an IR expansion for the fields given by:
\begin{eqnarray} \label{eqt:dilatonDyon}
f(r) &=& \frac{1}{r^2}  \sum_{n=0}^{\infty} \frac{f_n}{r^n}  \ , \quad g(r) = \frac{1}{r^2} \sum_{n=0}^{\infty} \frac{g_n}{r^n}  \ , \quad a(r) =  \sum_{n=0}^{\infty} \frac{a_n}{r^n} \ , \quad h(r) = \frac{1}{r} \sum_{n=0}^{\infty} \frac{h_n}{r^n}  \ , \quad \Phi(r) =  \sum_{n=0}^{\infty} \frac{\Phi_n}{r^n} \ ,  \nonumber \\
g_0 &=&  \frac{e^{2 \Phi_0 / \sqrt{3}}}{3 \left( 1 + e^{4 \Phi_0/\sqrt{3}} \right)} \ , \quad  \frac{h_0}{\sqrt{f_0}} = \frac{1}{\sqrt{ e^{2 \Phi_0/\sqrt{3}} + e^{2 \sqrt{3} \Phi_0}}}  \ , \quad \tilde{B}^2 = 3 a_0^2  \ .
\end{eqnarray}
The coefficients $f_n$ are completely undetermined as well as $\Phi_0$, $a_0$, and $a_1$.   In the IR, there are three free parameters: $a_1/a_0$, $\Phi_0$, and the function $f(r)$ (subject to obeying the right asymptotics).  This translates to a choice of the magnetic field, the chemical potential, and the source for the dilaton in the dual field theory.  Note that since we are studying the extremal case, the electric charge behind the horizon is fixed by the magnetic field:
\begin{equation}
\lim_{r \to \infty} Z(\Phi) (\ast F )_{xy} = e^{2 \Phi_0 / \sqrt{3}} \frac{a_0 h_0}{\sqrt{g_0 f_0}} = \tilde{B} \ .
\end{equation}
We now turn on the pressure, charge density, energy density due to the electron star and investigate the various gravitational solutions we can construct.
%
\section{Solutions with Stars}
%
\subsection{Mesonic Phase: Star in the Infra--Red}
\begin{figure}[ht] %
   \centering
   \includegraphics[width=3in]{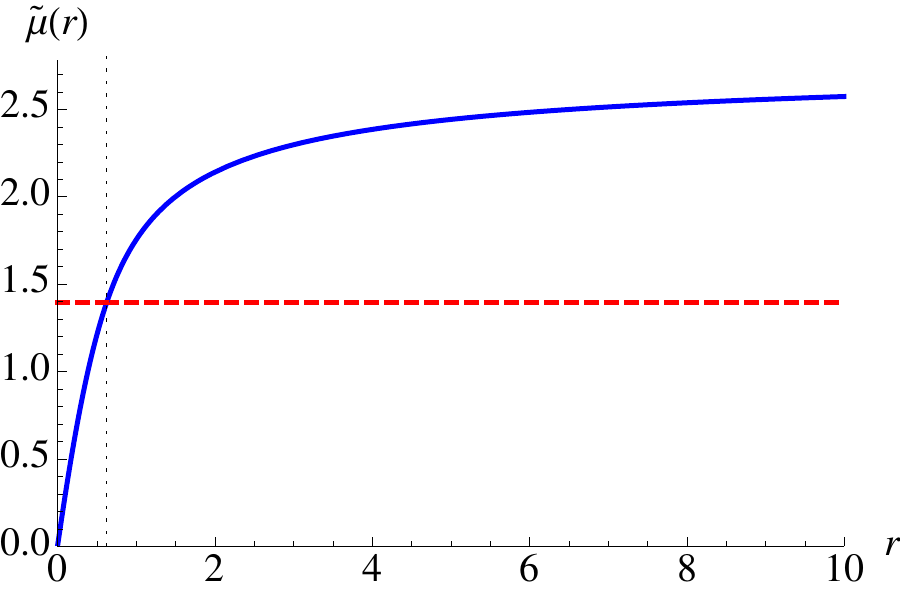} 
   \caption{\small Solution for the local chemical potential (solid blue curve) of a star in the IR using $f(r) = r^{-2}$ with $\tilde{\beta} = 1$, $\tilde{m} = 3 \cdot 3^{1/4} / (2 (2 \tilde{\beta})^{1/2})$, $\Phi_0 = - \sqrt{3}$.  The red dashed line is the value of $\tilde{m}$ and the black dotted line is the position where the star ends.  Corresponds to a solution with $\tilde{B} = 6.63$, $\mu = 3.08$, $\rho = 0.53$, $\phi_1 = -17.58$, $\phi_2 = 177.36$.}
   \label{fig:starIR}
\end{figure}

%
Consider an IR expansion for the fields:
\begin{eqnarray}
f(r) &=& \frac{1}{r^2}  \left( \sum_{n=0}^{\infty} \frac{f_n}{r^{2n/3}} \right) \  , \quad  g(r) = \frac{1}{r^{8/3}} \left(\sum_{n=0}^{\infty} \frac{g_n}{r^{2n/3}} \right) \ , \quad a(r) =  \frac{1}{r^{2/3}} \left(\sum_{n=0}^{\infty} \frac{a_n}{r^{2n/3}} \right) \ , \nonumber \\
\Phi(r) &=& - \frac{\sqrt{3}}{3} \ln r + \left(\sum_{n=0}^{\infty} \frac{\Phi_n}{r^{2n/3}} \right) \ , \quad h(r) = \frac{ 1}{r} \left( \sum_{n=0}^{\infty} \frac{h_n}{r^{2 n/3}} \right)  \ .
\end{eqnarray}
This solution is a natural extension of the purely magnetic dilaton-black hole reviewed in the previous section, and the leading order behavior is not changed from the pure magnetic dilatonic black hole except that the introduction of the star now changes the power of the expansion as well as turns on the gauge field $A_t$:
\begin{equation}
\tilde{B}^2 = \frac{3}{2} e^{-4 \Phi_0/\sqrt{3}} a_0^2  \ , \quad g_0 = \frac{16}{27} e^{2 \Phi_0/\sqrt{3}} \ , \quad \frac{h_0}{\sqrt{f_0}} = \frac{16}{81} \tilde{\beta} \left( \frac{h_0^2}{f_0} - \tilde{m}^2 \right)^{3/2} \ .
\end{equation}
There is no charge behind the horizon since $\lim_{r \to \infty} Z(\Phi) (\ast F )_{xy} \propto r^{-1/3}$. As in the sourceless case, the coefficients $f_n$ are undetermined by the equations of motion.  We can then solve for all remaining coefficients explicitly in terms of $a_0$, $f_n$, and $\Phi_0$ order by order as an expansion in the~IR.  We show an example of a solution in figure \ref{fig:starIR}.
%
\subsection{Partially Fractionalized Phases: Star outside Horizon}
\begin{figure}[ht]
\begin{center}
\subfigure[Star ends at horizon]{\includegraphics[width=3.0in]{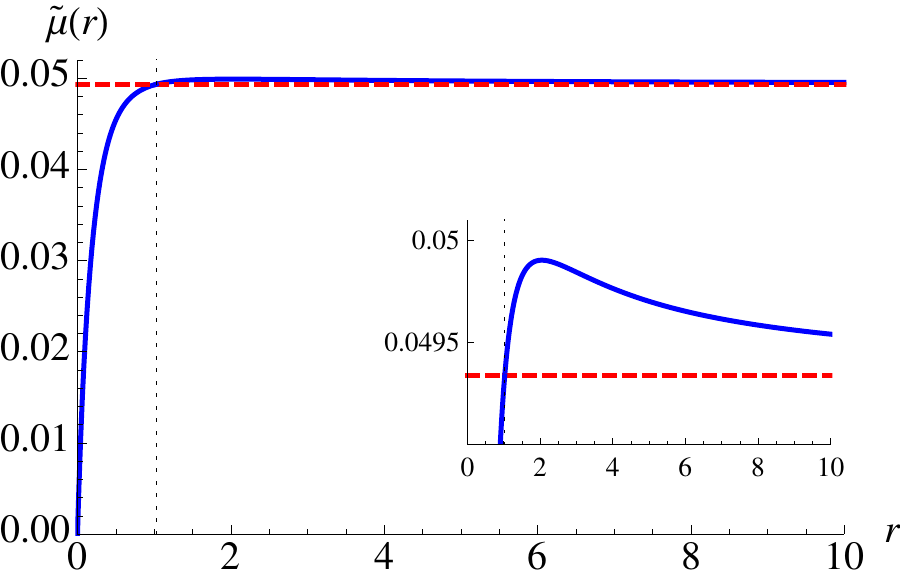} \label{fig:Fractionalizeda}} \hspace{0.5cm}
\subfigure[Star finite distance from horizon]{\includegraphics[width=3.0in]{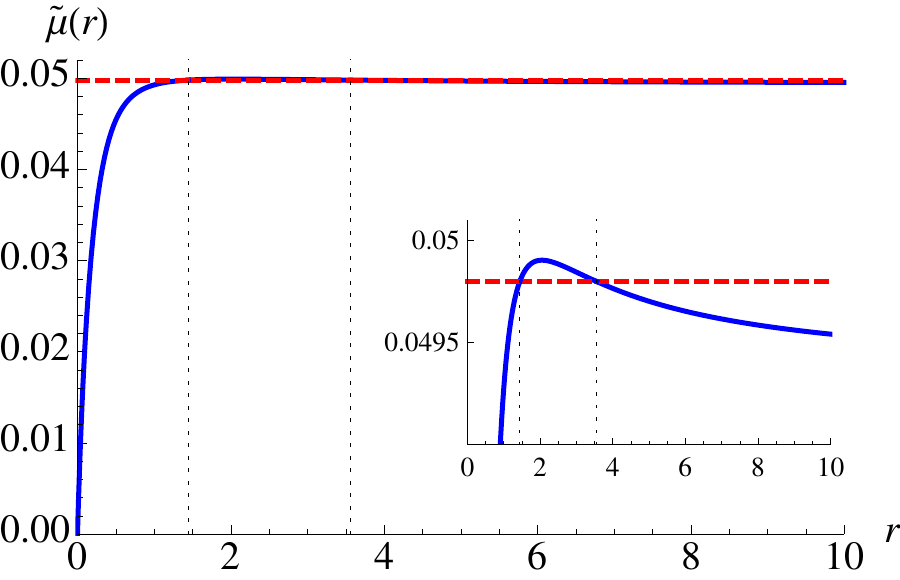} \label{fig:Fractionalizedb}}   
\end{center}
\caption{\small The local chemical potential (solid blue curve) for the partially fractionalized phases with (a) the star ending at the horizon and (b) the star a finite distance from the horizon.  The dashed red line is the mass $\tilde{m}$ and the dot dashed black line is where the star starts/ends.  For both $f(r) = r^{-2}, \tilde{\beta} = 10, \Phi_0 = \sqrt{3}$.  (a) $\tilde{m} =  \frac{1}{\sqrt{ e^{2 \Phi_0/\sqrt{3}} + e^{2 \sqrt{3}  \Phi_0}}} , \tilde{B} = 1.28, \mu = 0.3, \rho = 1.28, \phi_1 = 7.49, \phi_2 = -30.86$. (b) $\tilde{m} = 0.0498 , \tilde{B} = 1.28, \mu = 0.3, \rho = 1.28, \phi_1 = 7.47, \phi_2 = -30.82$.  In these examples, the star has a negligible effect on the overall charge of the system.}
   \label{fig:Fractionalized}
\end{figure}

%
For the case of having no star in the IR with an electric charge behind the horizon, there are two cases to consider: the case where the star is a finite distance away from the horizon, and the case where it is not.  We first consider the latter.  The IR expansion is given by:
\begin{eqnarray}
f(r) &=& \frac{1}{r^2}\left( \sum_{n=0}^{\infty} \frac{f_n}{r^{n/2}} \right) \ , \quad g(r) = \frac{1}{r^2} \sum_{n=0}^{\infty} \frac{g_n}{r^{n/2}}  \ , \quad a(r) =  \sum_{n=0}^{\infty} \frac{a_n}{r^{n/2}} \ , \quad h(r) = \frac{1}{r} \sum_{n=0}^{\infty} \frac{h_n}{r^{n/2}}    \ , \nonumber  \\
 \Phi(r) &=& \sum_{n=0}^{\infty} \frac{\Phi_n}{r^{n/2}}  \ , \quad B^2 = 3 a_0^2  \ , \quad  g_0 =  \frac{e^{2 \Phi_0 / \sqrt{3}}}{3 \left( 1 + e^{4 \Phi_0/\sqrt{3}} \right)} \ , \quad  \frac{h_0}{\sqrt{f_0}} = \pm  \frac{1}{\sqrt{ e^{2 \Phi_0/\sqrt{3}} + e^{2 \sqrt{3}  \Phi_0}}}  \  , \nonumber \\
 \tilde{m} &=&  \frac{1}{\sqrt{ e^{2 \Phi_0/\sqrt{3}} + e^{2 \sqrt{3}  \Phi_0}}}  \ , \quad f_1 = 0 =  h_1  = 0= g_1  = 0 = a_{1} = 0 \ . 
\end{eqnarray}
All higher $n \geq 2$ coefficients are non--zero with $a_2/a_0$ and $f_{n\geq 2}$ as free parameters.  The leading IR behavior is unchanged from the dyon, except that we find that the value of $\Phi_0$ is fixed by the mass~$\tilde{m}$, and the expansion is now in terms of  fractional powers.  An example is shown in figure \ref{fig:Fractionalizeda}.

The local chemical potential at the horizon is equal to $\tilde{m}$ meaning that at the horizon we have no star.  However, we find that $\tilde{\mu}$ in the IR evolves as:
\begin{equation}
\frac{h(r)}{\sqrt{f(r)}} = \tilde{m} + \frac{1}{r} \left( \frac{e^{-\Phi_0/\sqrt{3}} \left( -5 - 8 e^{4 \Phi_0/\sqrt{3}} + e^{8 \Phi_0/\sqrt{3}} \right) }{18 \left( 1 + e^{4 \Phi_0}{\sqrt{3}} \right)^{5/2}} \frac{a_2}{a_0} \right) + O(r^{-2}) \ .
\end{equation}
The second term in the expansion appears in the value of other coefficients to some fractional power, and therefore is required to be positive 
\begin{equation}
\left( -5 - 8 e^{4 \Phi_0/\sqrt{3}} + e^{8 \Phi_0/\sqrt{3}} \right) \frac{a_2}{a_0} > 0 \ .
\end{equation}
We are also required to take $a_2 > 0$ in order to get an asymptotically AdS solution (this observation is found numerically). Therefore, since for star formation we need the chemical potential to increase, we have a condition on $\Phi_0$:
\begin{equation}
\Phi_0 > \frac{\sqrt{3}}{4} \ln \left( 4 + \sqrt{21} \right) \equiv \Phi_{c} \ .
\end{equation}
For $\Phi_0 < \Phi_c$, the chemical potential decreases monotonically from its value at the horizon and does not allow a star to form.  For $\Phi_0 = \Phi_c$,  the solution is AdS$_2 \times R^2$ everywhere (if we choose $f(r)~=~r^{-2}$).  This solution therefore separates the partially fractionalized from the fully fractionalized solution for this class of star backgrounds. 

There is also the possibility of having a star that is a finite distance from the horizon.  In this case, the IR expansion would be that of the dilaton-dyon in equation \reef{eqt:dilatonDyon}.  It requires that  
\begin{equation}
h_0 < \tilde{m} < \tilde{m}_{max} \ , \quad \Phi_0 > \Phi_c \ ,
\end{equation}
where $\tilde{m}_{max}$ is the maximum mass after which the chemical potential never gets high enough to populate the star. An example is shown in figure \ref{fig:Fractionalizedb}.
%

\section{Concluding Remarks}
%
We have constructed three types of zero temperature electron star solutions with background magnetic field. Note that, including the solution without a star, our work does not address which of the four solution types is the thermodynamically preferred state for fixed chemical potential, dilaton source, and magnetic field.  this is for future work. For the case of vanishing magnetic field, such a  phase diagram at zero temperature was constructed \cite{Hartnoll:2011pp}, where it was observed that a Lifschitz fixed point separates the mesonic and partially fractionalized phases when the phase transition is continuous (the Lifschitz point is avoided in the case of the first order transition), and a third order transition separates the partially fractionalized phase from the fully fractionalized phase.  Here, we have observed that an AdS$_2\times \mathbb{R}^2$ geometry appears to separate the fractionalized phase from the fully fractionalized phase at finite magnetic field, and it would be interesting to study how this point affects the phase transition.  

We note the absence of any Lifshitz solutions in our presentation of the various phases so far.  In ref.~\cite{Hartnoll:2011pp}, the Lifshitz solution plays the important role of mediating between the mesonic phase (with no horizon) from the partially fractionalized phase (with an electrically charge horizon) at zero magnetic field.  However, as indicated earlier, in the presence of a magnetic field, all the phases we study involve a horizon behind which the magnetic charge hides, and the transition from the mesonic phase to the fractionalized phase is a transition between an electrically neutral horizon and an electrically charged horizon.  This transition may be softer, which might explain our inability to find a mediating solution.  This picture is of course incomplete without an action calculation to study the exact nature of the transition between the two phases. (Note that  since the dilaton diverges on the horizon, some care may be needed in the treatment of these solutions, perhaps by exploiting their M--theory uplift mentioned in the introduction.)

Another obvious generalization of our work would be to construct finite temperature solutions as was done for the electron star at zero magnetic field \cite{Puletti:2010de,Hartnoll:2010ik}.  Finally, it would be interesting to explore  the transport properties of our systems, employing  fluctuation analyses and computing the appropriate Green's functions. We will leave these matters for another day.

\section*{Acknowledgements}
TA, CVJ, and SM would like to thank the  US Department of Energy for support under grant DE-FG03-84ER-40168.  TA is also supported by the USC Dornsife College of Letters, Arts and Sciences. We thank Kristan Jensen, Joe Polchinski, and Herman Verlinde  for  comments.

\providecommand{\href}[2]{#2}\begingroup\raggedright\endgroup


\end{document}